\documentclass[a4paper,fleqn,usenatbib]{mnras}

\usepackage[T1]{fontenc}
\usepackage{ae,aecompl}
\usepackage{epsf,rotating}
\usepackage{amsmath}
\usepackage[]{hyperref}
\usepackage{amssymb}
\usepackage{color}
\usepackage{verbatimbox}
\usepackage{bm}
\usepackage{bigfoot}
\usepackage{cleveref}
\usepackage{url,hyperref}
\newcommand{\msolar}{\;{\rm M}_{\odot}}
\newcommand{\fesc}{f_{\mathrm{esc}}}
\newcommand{\fescs}{f_{\mathrm{esc,\star}}}
\newcommand{\fesca}{f_{\mathrm{esc,AGN}}}

\newcommand{\rion}{{R_{\rm ion}}}
\newcommand{\ragn}{{R_{\rm AGN}}}
\newcommand{\rst}{{R_{\rm \star}}}
\newcommand{\rrec}{{R_{\rm rec}}}

\title[EoR Models Classifier]{Identifying Reionization Sources from 21cm Maps using Convolutional Neural Networks}

\author[S. Hassan et al.]{
\parbox[t]{\textwidth}{\vspace{-1cm}
Sultan Hassan$^{1,2}$\thanks{SKA Fellow}\thanks{E-mail: sultanier@gmail.com}, Adrian Liu$^{3,4}$\thanks{Hubble Fellow}, Saul Kohn$^5$, Paul La Plante$^5$}
\\
\\$^1$ University of the Western Cape, Bellville, Cape Town, 7535, South Africa
\\$^2$ New Mexico State University, Las Cruces, NM 88003, United States
\\$^3$ Department of Astronomy and Radio Astronomy Laboratory, University of California Berkeley, Berkeley, CA 94720, United States
\\$^4$ Department of Physics and McGill Space Institute, McGill University, Montreal QC H3A 2T8, Canada
\\$^5$ Department of Physics and Astronomy, University of Pennsylvania, Philadelphia, PA 19104, United States
}
\date{Accepted XXX. Received YYY; in original form ZZZ}

\pubyear{2017}

\begin{document}
\label{firstpage}
\pagerange{\pageref{firstpage}--\pageref{lastpage}}
\maketitle

\begin{abstract}
Active Galactic Nuclei (AGN) and star-forming galaxies are leading candidates for being the luminous sources that reionized our Universe. Next-generation 21cm surveys are promising to break degeneracies between a broad range of reionization models, hence revealing the nature of the source population. While many current efforts are focused on a measurement of the 21cm power spectrum, some surveys will also image the 21cm field during reionization. This provides further information with which to determine the nature of reionizing sources. We create a Convolutional Neural Network (CNN) that is efficiently able to distinguish between 21cm maps that are produced by AGN versus galaxies scenarios with an accuracy of 92-100\%, depending on redshift and neutral fraction range. An exception to this is when our Universe is highly ionized, since the source models give near-identical 21cm maps in that case. When adding thermal noise from typical 21cm experiments, the classification accuracy depends strongly on the effectiveness of foreground removal. Our results show that if foregrounds can be removed reasonably well,  SKA, HERA and LOFAR should be able to discriminate between source models with greater accuracy at a fixed redshift. Only future SKA 21cm surveys are promising to break the degeneracies in the power spectral analysis.

\end{abstract}




\section{Introduction}
Low-frequency radio interferometer experiments, such as the Low Frequency Array~\citep[LOFAR;][]{van13}, the Precision Array for Probing the Epoch of Reionization~\citep[PAPER;][]{par10}, the Hydrogen Epoch of Reionization Array~\citep[HERA;][]{debo17}, the Murchison Wide field Array~\citep[MWA;][]{bow13,ting13}, the Giant Metre-wave Radio Telescope~\citep[GMRT;][]{paci11}, and the Square Kilometer Array~\citep[SKA;][]{mel15}, are promising instruments for detecting reionization on cosmological scales in the near future. These growing observational efforts 
motivate the development of statistical tools for efficiently extracting astrophysical and cosmological information.

The nature of sources driving cosmic reionization remains a debated topic. While many studies~\citep[e.g.][]{shgi87,hop07,hama12,hassan18} conclude that reionization is mainly driven by star-forming galaxies, there remain some discrepancies that might be resolved by allowing Active Galactic Nuclei (AGN) to reionize the Universe. These discrepancies include the flat slope of ionising emissivity measurements~\citep{bec13}, the early and extended timing of Helium reionization~\citep{wor16}, and large scale intergalactic opacity fluctuations~\citep{bec15}. In addition, the \cite{gia15} high-redshift AGN measurements indicate a flatter slope at the faint end of the luminosity function, hence implying the presence of sufficient AGN to complete reionization as opposed to what was previously believed.

In~\citet{hassan18}, we studied the role of galaxies versus AGN during reionization using an improved semi-numerical simulation. This simulation implements realistic and physically motivated recipes to include ionizing photons from galaxies, following the~\citet{fin15} radiative transfer simulations, and from AGN, following the~\citet{gia15} observations and a fixed Quasar Halo Occupancy Distribution. We found that AGN-dominated scenarios are unable to simultaneously match current reionization constraints, namely the \cite{planck16} optical depth, ionising emissivity measurement by \cite{bec13}, and the \cite{fan06} neutral fraction constraints at the end of reionization. This shows that AGN are highly unlikely to be the sole drivers of cosmic reionization. However, a model with equal photon contributions from AGN and galaxies is still allowed given the large uncertainties of current observational constraints. The broad range of theoretical models allowed by current reionization observations is expected to narrow when data from 21cm surveys become available. We found in~\citet{hassan18} that power spectrum measurements from LOFAR, HERA and SKA could potentially discriminate between galaxies-only models and those with AGN contributions.
This is because the latter produce larger ionized bubbles, and hence boost the 21cm power spectrum by a factor of $\sim 2$ as compared to the former.

Future 21cm observations are also expected to provide huge datasets of large-scale 21cm images that will encode more information than just the 21cm power spectrum. These images might provide an alternative approach to constrain the contribution of different reionizing sources.  Here we assess the viability of using Convolutional Neural Networks (CNNs) to discriminate between AGN and galaxy models. This approach has been successfully employed in astronomy and cosmology to perform regression~\citep[e.g.][]{Gupta18,gil18} and classification~\citep[e.g.][]{ara18,company18,sch18}.

In this paper, we first examine the CNN capability to discriminate between these models at several redshifts as a function of neutral fraction when the simulations are free of instrumental effects. We then include real-world survey effects, adding realistic instrumental noise and foreground mitigation cuts in order to test the CNN performance on datasets as they will be seen by the SKA, HERA, and LOFAR surveys. We finally consider the case where our different reionization models share similar power spectra. This is essentially a test of our CNN's ability to break degeneracies inherent in a power spectral analysis. 

This paper is organized as follows: Section~\ref{sec:sim} provides a brief summary of the simulations used for this study, a description of our two reionization models (galaxies, AGN), and details of our instrument simulations. We introduce Convolutional Neutral Networks and describe the classifier architecture in Section~\ref{sec:classifier}.
We describe our training datasets in Section~\ref{sec:dataset}, present the results in Section~\ref{sec:results} and summarize our conclusions in Section~\ref{sec:con}.

\section{Simulations}\label{sec:sim}
We use the Time-integrated version of our semi-numerical code {\sc SimFast21}~\citep{san10} that has been developed in~\citet{hassan17}. We describe here the general components of our simulation and refer to~\citet{san10} for the full description of the algorithm and to~\cite{hassan16,hassan17} for more details on the Time-integrated version development.

We first generate the dark matter density field using a Monte-Carlo
Gaussian approach in the linear regime. We then dynamically evolve 
the linear density field into a non-linear field by applying 
the~\citet{zeldovich70} approximation. Using the underlying non-linear density
realizations, the dark matter halos are identified using the well known 
excursion set formalism~\citep[ESF,][]{prsc74,bond99} in which a region $x$
is assumed to collapse into halos, if its mean overdensity $\delta(x)$ is  higher than an overdensity threshold of 
$\delta_{c}(z) \sim 1.68/D(z)$, where $D(z)$ is the linear growth factor. Around each point in the density grid, the mean overdensity is obtained through a sequence of top-hat filters, starting with the largest possible size (the box size) and decreasing down to the smallest size (the cell size). The minimum halo mass is set to 10$^{8}\, \msolar$, which corresponds to the hydrogen cooling limit during reionization.
In the Time-integrated model, the ionised regions are
identified using a similar ESF that
is based on comparing the time-integrated ionisation rate $\rion$
with that of the recombination rate $\rrec$ and the local neutral hydrogen
density within each spherical volume specified by the ESF.  Regions are flagged as ionised if
\begin{equation}
\label{eq:ion_con}
\int\fesc \rion \,\, dt \geq \int x_{\rm HII}\, \rrec\,\, dt +  (1 - x_{\rm HII})\, N_{\rm H}\, ,
\end{equation}
where $\fesc$ is the photon escape fraction, $x_{\rm HII}$ is the
ionised fraction, and $N_{\rm H}$ is the total number of hydrogen
atoms. Equation~\eqref{eq:ion_con} represents the ionisation condition
used in the Time-integrated model. The recombination field $\rrec$ is modelled
using a derived parametrization taken from a high-resolution radiative simulation
~\citep{fin15} to account for the small scale inhomogeneity below our cell size and
clumping factor effects. The recombination rate, $\rrec$, is parametrized as a function of overdensity $\Delta$
and redshift $z$  as follows:
\begin{equation}\label{eq:rrec}
\frac{\rrec}{V} =  A_{\rm rec}(1+ z)^{D_{\rm rec}}  \left[\frac{\left( \Delta/B_{\rm rec} \right)^{C_{\rm rec}}}{1+ \left( \Delta/B_{\rm rec} \right)^{C_{\rm rec}} } \right]^{4} \, , 
\end{equation}
where $A_{\rm rec} = 9.85 \times 10^{-24} $cm$^{-3}$s$^{-1}$ (proper
units), $B_{\rm rec}=1.76 $, $C_{\rm rec}= 0.82$, $D_{\rm rec}=5.07$,  and $V$ 
refers to the cell size. 
We refer to~\citet{hassan16} for more details on the derivation of $\rrec$ and the
effect of this parameter on reionization observables. 

The key difference between the galaxy and AGN models is in their prescriptions for computing $\rion$. We dub this variable $\rst$ for the galaxy model and $\ragn$ for the AGN model.

\subsection{Galaxies}
Our parameterization for the ionizing photon output from galaxies $\rst$ is motivated by a combination of
radiative transfer simulations~\citep{fin15} and larger hydrodynamic
simulations~\citep{dav13} that have been calibrated to match wide
range of observations. In particular, $\rst$ is parametrized as a function of 
halo mass $M_{\rm h}$ and redshift $z$ as follows:
\begin{equation}\label{eq:nion}
\frac{\rst}{M_{\rm h}} =  A_{\rm \star}\, (1 + z)^{D_{\rm \star}} \, ( M_{\rm h}/B_{\rm \star} )^{C_{\rm \star}} \, \exp\left( -( B_{\rm \star}/M_{\rm h})^{3} \right),
\end{equation}
where $A_{\rm \star} =1.08\times 10^{40}  \msolar^{-1} $s$^{-1}$,
$B_{\rm \star} = 9.51\times 10^{7}\msolar$, $C_{\rm \star} = 0.41$ and
$D_{\rm \star} = 2.28$.
This ionization rate $\rst$ is computed directly from the star formation rate (SFR)
 of galaxies in these hydrodynamic simulations. This $\rst$ parametrization
has been dervied fully in~\cite{hassan16}.  The emissivity amplitude $A_{\rm \star}$ here scales the ionization rate $\rst$ by the same amount for every ionizing photon source and adjusts the total ionizing emissivity evolution in redshift. The $C_{\rm \star}$, in turns, scales the ionization rates super-linearly with halo mass, insuring that more massive halos emit more photons than the less massive ones, and hence generating larger ionized bubbles. This is a key parameter to control the ionized bubble sizes. The $B_{\rm \star}$ represents the halo mass quenching scale for the turn over at the low halo masses and faint ionizing sources. As previously considered in~\citet{hassan17}, we keep all other 
parameters fixed and vary the photon escape fraction $\fescs$, the emissivity amplitude $A_{\rm \star}$ and
the emissivity-halo mass power dependence index $C_{\rm \star}$ to produce different ionizing emissivities $\dot{N}_{\rm \star}$, for our training samples, as follows:
\begin{equation}\label{eq:N_star}
\dot{N}_{\rm \star}  = \fescs \rst = {\rm const.}\,\, \fescs \, A_{\rm \star}\, M_{\rm h}^{C_{\rm \star}+1}\,,
\end{equation} 
where the proportionality constant is
\begin{equation}
{\rm const.} = (1 + z)^{D_{\rm \star}}\,B_{\rm \star}^{- C_{\rm \star}}\, \exp\left( -( B_{\rm \star}/M_{\rm h})^{3} \right).
\end{equation}

\subsection{AGN}
We compute the ionizing photon output from AGN using the strong correlations seen in local Universe observations~\citep{fer02,tre02} between black hole masses $M_{\rm bh}$ and circular velocities $v_{\rm cir}$ of their host halos. This correlation is written as
\begin{equation}\label{eq:mbh}
\frac{M_{\rm bh}}{M_{\odot}} = A_{\rm AGN} \, \left(\frac{v_{\rm cir}}{159.4\,{\rm km\,s^{-1}}} \right)^{5}\, , 
\end{equation}
where the constant $A_{\rm AGN}$ may be regarded as the black hole formation efficiency,
which is a free parameter in the AGN source model besides $\fesca$. Assuming the Eddington
luminosity in the B-band ~\citep{cho05} and a power law spectral energy distribution, the AGN ionization rate is given by:
\begin{equation}\label{eq:ragn}
\ragn = \int_{\nu_{912}}^{\infty} \, \frac{L_{912}}{h\nu} \left(\frac{\nu}{\nu_{912}}\right)^{-1.57} \, d\nu =  \frac{L_{912}}{1.57\, h},
\end{equation}
where $h$ is Planck's constant and 
\begin{equation*}\label{eq:l912}
\frac{L_{912}}{\rm ergs\, s^{-1}\, Hz^{-1}} = 10^{18.05}\, \frac{L_{B}}{L_{\odot,B}}\,\,\textnormal{and}\,\,  \frac{L_{B}}{L_{\odot,B}} = 5.7\times 10^{3}\, \frac{M_{\rm bh}}{M_{\odot}}\, ,
\end{equation*}
where the $L_{912}$ and $\nu_{912}$ are the luminosity and frequency at the Lyman limit, corresponding to 912  \AA.
To study the effect of massive and less massive AGN, it is reasonable to assume a non-linear relation between $\ragn$ and $M_{\rm bh}$, due to the fact that more massive AGN have stronger feedback effects than the smaller ones, which implies more photons are being produced. This is, in principle, similar to our Galaxies source model in which $\rst$ depends non-linearly on $M_{\rm h}$ through $C_{\rm \star}$. Thus, we introduce a new parameter $C_{\rm AGN}$ in the above relations. This leads to an AGN model with three free parameters:  AGN photon escape fraction $\fesca$, ionising emissivity-black hole mass dependence (the non-linear scale index) $C_{\rm AGN}$, and ionising emissivity amplitude $A_{\rm AGN}$ (equivalent to the black hole formation efficiency). We keep all other parameters fixed and vary these parameters to generate different ionizing emissivities $\dot{N}_{\rm AGN}$, for our training datasets, using:  
\begin{equation}\label{eq:N_agn}
\dot{N}_{\rm AGN}  = \fesca \ragn = {\rm const.}\,\fesca \,A_{\rm AGN}\,M_{\rm bh}^{C_{\rm AGN}+1}.
\end{equation}
where the proportionality constant is ${\rm const.} = 6.15 \times 10^{47}$. 
This shows how the AGN ionizing emissivity is related to the host halo mass. We next use the~\citet{gia15} luminosity function at $z\sim 6$ to find the corresponding AGN duty cycle for each halo mass bin, which is the ratio between the number of AGN to the number of possible host halos. This ratio is alternatively called the Quasar Halo Occupancy distribution (QHOD). The QHOD then provides a probability for a particular host halos to have an AGN that is actually active, and we use the QHOD to randomly designate a given halo as one that produces ionizing photons according the prescription outlined above. Instead of using a fixed luminosity function, we use a fixed QHOD at higher redshifts to find the number of AGN.  This is motivated by the fact that QHOD doesn't evolve strongly with redshift as the AGN luminosity functions do. We defer to~\citet{hassan18} for the full details of our AGN source model prescription. 
\begin{figure*}
\centering
\setlength{\epsfxsize}{0.5\textwidth}
\centerline{\includegraphics[scale=0.65]{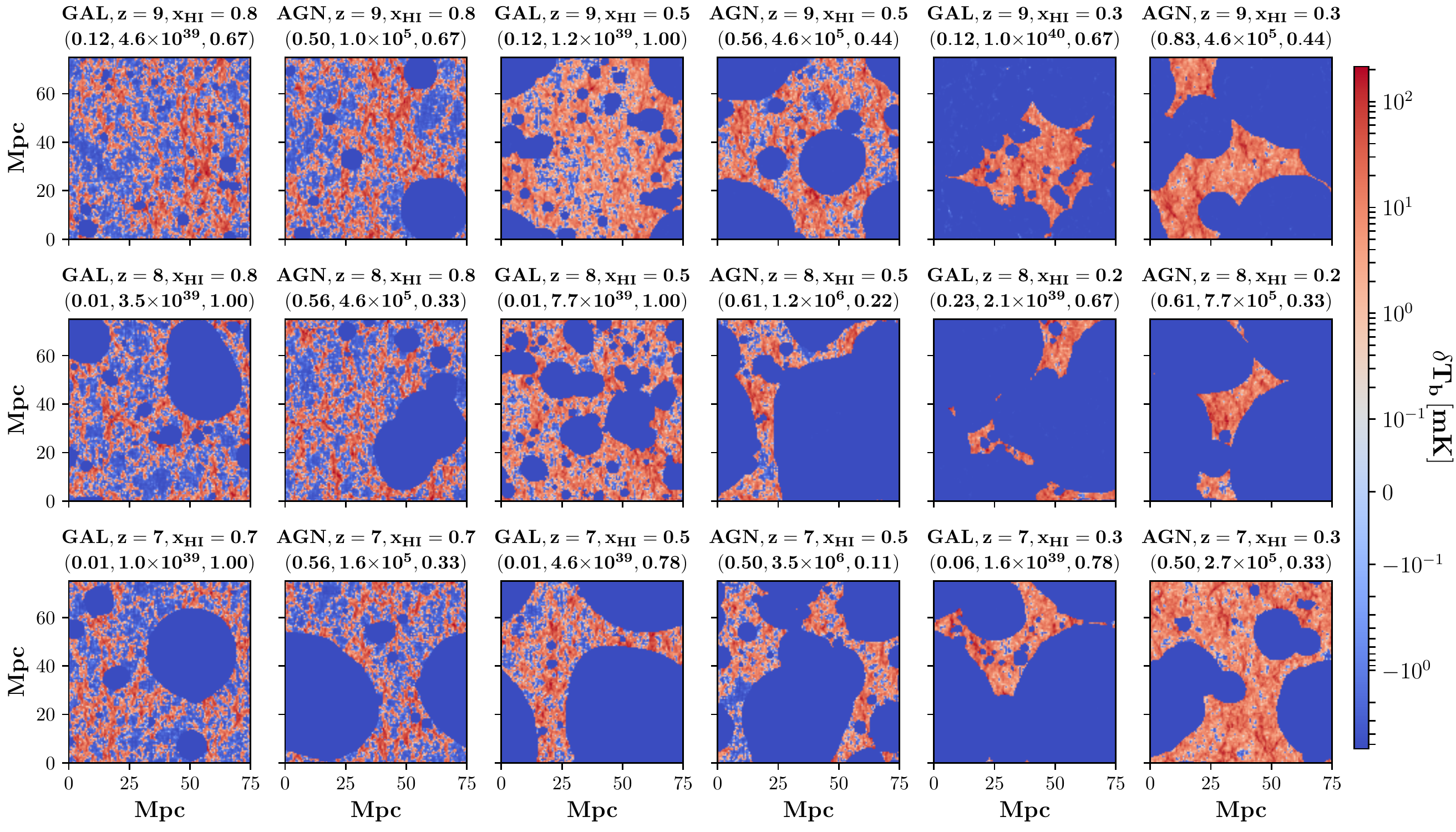}}
\caption{ Several 21cm images, each with a thickness of a single cell size (0.535 Mpc), as produced by Galaxies (odd columns) and AGN (even columns) at $z$= 9, 8, 7 at $x_{\rm HI} \sim$ 0.75, 0.5, 0.25. We quote in subtitles the astrophysical parameters used for Galaxies ($\fescs, A_{\rm \star}, C_{\rm \star}$) and for AGN ($\fesca,  A_{\rm AGN}, C_{\rm AGN}$). Regardless these parameters degeneracy, the AGN-only models tend to produce larger ionized bubbles whereas Galaxies-only models produce more small scale ionized bubbles. }
\label{fig:some_maps}
\end{figure*} 

In Figure~\ref{fig:some_maps},  we show different 21cm images for Galaxies and AGN models at $z= 9$, 8, 7 at different stages of reionization when x$_{\rm HI} \sim$ 0.27, 0.5, 0.25.  Each image is extracted from an 21cm box that is generated from a different realization of the underlying density field (different seed) and different set of astrophysical parameters. These parameters are quotes above each image, in the format ($\fescs, A_{\rm \star}, C_{\rm \star}$) for Galaxies and ($\fesca,  A_{\rm AGN}, C_{\rm AGN}$) for AGN. These images are randomly selected from our training datasets, whose assembly we explain in greater detail in Section \ref{sec:dataset}. We see that Galaxies scenarios (odd columns) have more small scale bubbles than AGN scenarios (even columns), which tend to have larger bubbles. As mentioned before, the size of ionised bubbles in each model is basically controlled by the emissivity's dependence on mass, namely $C_{\rm \star}$ and $C_{\rm AGN}$ for Galaxies and AGN, respectively.  In other words, $C_{\rm \star}$ and $C_{\rm AGN}$  parameters quantify the models' departure from the linear to non-linear dependence on mass. However, the degeneracy among these parameters (see eq.~\ref{eq:N_agn} and~\ref{eq:N_star}) and the variation in the underlying density field for each realization prevent direct comparison between these maps to isolate the impact of each individual parameter. In~\citet{hassan18} and by using the same density field realization, we have shown that the AGN models (with $C_{\rm AGN}=0.0$ corresponding to a linear relation between $\dot{N}_{\rm AGN}$ and $M_{\rm bh}$) still produce larger ionized bubbles than Galaxies models (with $C_{\rm \star}=0.44$ corresponding to non-linear relation between $\dot{N}_{\rm \star}$ and $M_{\rm h}$). Regardless these parameters degeneracy, we can see some differences as follows: 
\begin{itemize}
\item Galaxies models require much larger $C_{\rm \star}$ value than AGN models use for $C_{\rm AGN}$ to produce larger ionized bubbles (see fig.~\ref{fig:some_maps} at z=8,7 at $x_{\rm HI}$=0.8,0.3, respectively). 
\item For Galaxies and AGN models that have the same $C_{\rm \star}$ and $C_{\rm AGN}$ values, AGN models will typically still produce larger ionised bubbles at same neutral fraction and redshift. (see fig.~\ref{fig:some_maps}  at $z= 9$ and $x_{\rm HI}$= 0.8).
\end{itemize}

Of the remaining parameters, the photon escape fraction $\fesc$ and the emissivity amplitude $A$ linearly increase the ionising emissivity by the same amount for every source, and hence they have a bigger effect on the reionization history than on the reionization topology.  Figure~\ref{fig:some_maps} displays a random sample from our training dataset for the purpose of showing the models' variations and parameters degeneracy which we hope to distinguish using the developed classifier.

To summarize, the main differences between our Galaxies and AGN models are:
\begin{itemize}
\item Both models have ionizing emissivities that scale super-linearly with halo mass, but the Galaxies model includes an exponential cut-off at low masses; see Equations~\eqref{eq:N_agn} and \eqref{eq:N_star}.
\item The AGN model accounts for redshift evolution following their host halos, whereas the Galaxies model has an extra redshift evolution factor which is important to match the ionizing emissivity evolution at higher redshifts as compared with observations (e.g.~\citealt{bec13} ionising emissivity measurements); see Equation~\eqref{eq:N_star}. 
\item In Galaxies model, the star-formation duty cycle is assumed to be unity, which means that all identified halos are forming stars and contributing to the total ionizing emissivity. On the other hand, AGN models implement a duty cycle following the~\citet{gia15} luminosity function, which means that not all halos host AGN and contribute to the photon production rate.
\end{itemize}

\subsection{21cm Instrument Simulations}\label{sec:noise}
We now describe our methodology for including realistic 21cm instrumental effects. Doing so requires accounting for the finite angular resolution of telescopes, the effects of foreground cleaning, and the presence of instrumental noise.

\begin{figure*}
\centering
\setlength{\epsfxsize}{0.5\textwidth}
\centerline{\includegraphics[scale=.5]{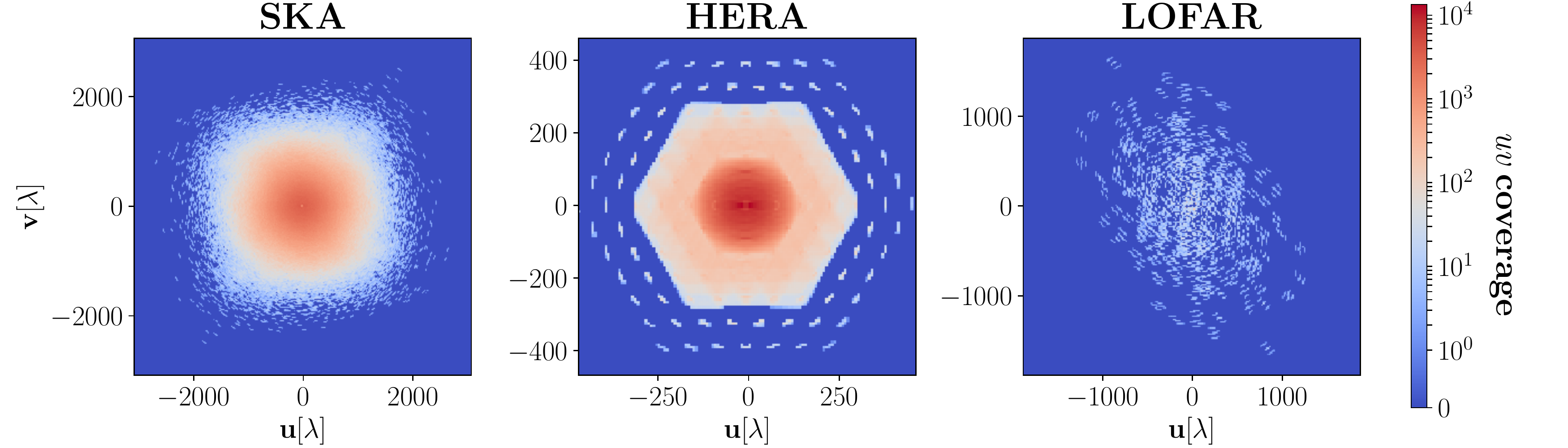}}
\caption{ The $uv$ coverage for the SKA, HERA and LOFAR arrays at $\nu$=157.8 MHz (z=8) using {\sc 21cmSense}. Zero coverage pixels (blue color)
correspond to non-measured signal modes which lie beyond the experiments' layout angular resolution.}
\label{fig:uv_map}
\end{figure*}
\subsubsection{Angular resolution}
We account for the finite angular resolution of an 21cm experiment by exploiting its detailed baseline distribution. The 21cm brightness of a batch in the sky along $x$ and $y$ coordinates can be converted in terms of baseline length in $u$ and $v$ coordinates. The comoving wavenumber in the directions perpendicular to our line of sight (i.e., the angular directions) is given by ${\bf k}_{\perp} = 2 \pi {\bf u}_{\perp}/ D_{c}$, where $D_{c}$ is the comoving distance to the observed 21cm batch at redshift $z$. The ${\bf k}_{\perp}$ and ${\bf u}_{\perp}$ here are expressed in terms of $k_{x}-k_{y}$ and $u-v$ components accordingly. In the $u-v$ plane, we compute the $uv$ coverage which represents the total number of baselines that observe a given $uv$ pixel using {\sc 21cmSense}\footnote{\url{https://github.com/jpober/21cmSense}}. In Figure~\ref{fig:uv_map}, we show the $uv$ coverage for our three 21cm array designs, namely the SKA, HERA and LOFAR at z=8 ($\nu=157.8$ MHz). The zero pixels (blue color) show the signal modes that lie beyond the experiments angular resolution. This shows the angular resolution limitation for each experiment. The SKA and HERA have higher $uv$ coverages (red color) as compared to the sparse baseline distribution in case of LOFAR, and hence we expect better classification performance for these experiments.

Our recipe for adjusting the 21cm brightness temperature images to account for an instrument's angular resolution is as follows:

\begin{itemize}
\item We compute the $uv$ coverage from the antenna distribution of the 21cm experiment in question at a given frequency (redshift). 
\item We convert the $u$ and $v$ coordinates into their corresponding $k_{x}$ and $k_{y}$ modes. 
\item We Fourier transform the 21cm image.
\item We zero out all Fourier modes corresponding to zero $uv$ coverage at $k_{x}$ and $k_{y}$, assuming that these modes are non-measured.
\item We inverse Fourier back to the real space to obtain the angular resolution limited 21cm image.
\end{itemize}

\begin{figure*}
\centering
\setlength{\epsfxsize}{0.5\textwidth}
\centerline{\includegraphics[scale=1.0]{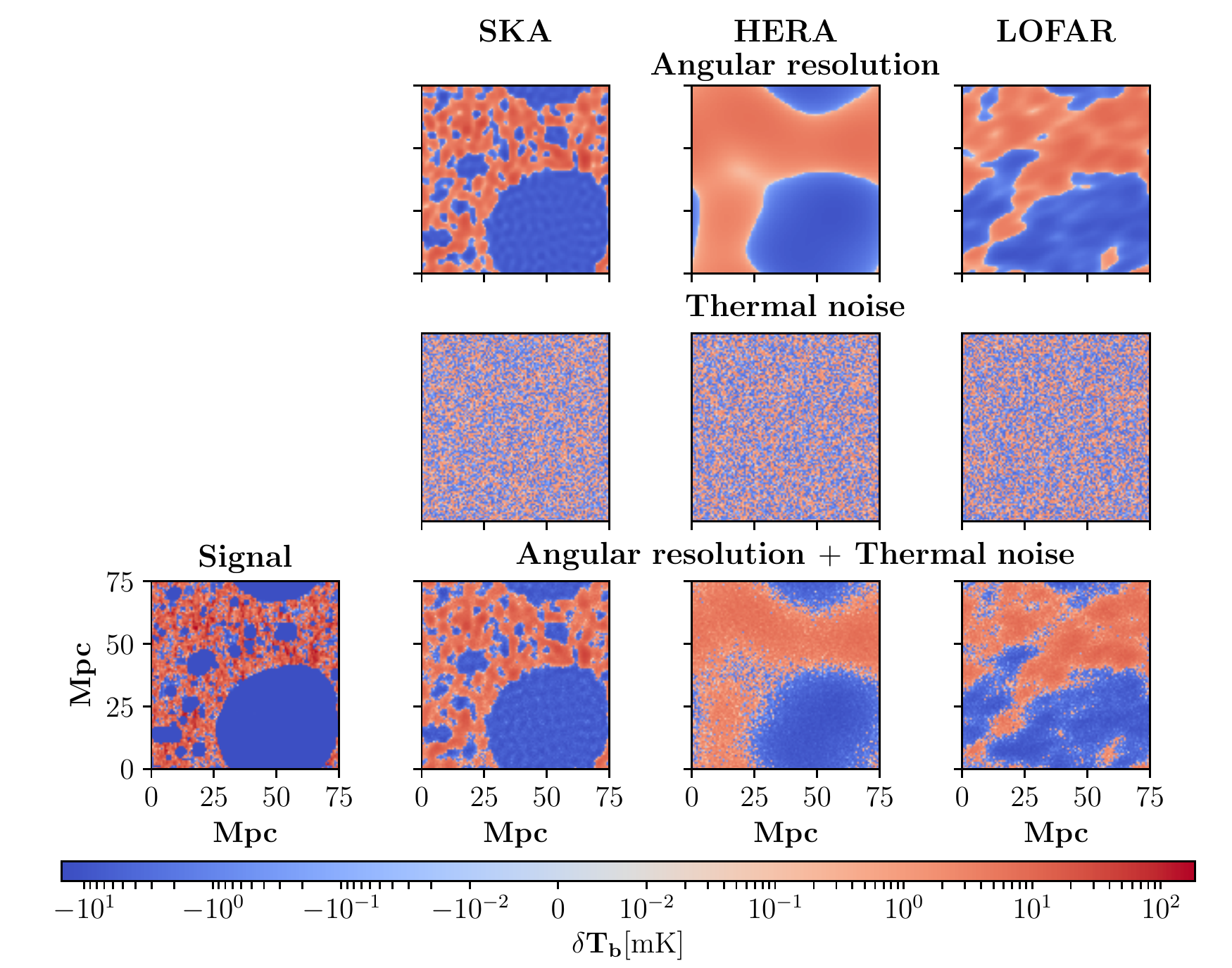}}
\caption{ A summary of the 21cm instrumental effect pipeline. {\sc Bottom left}: a 21cm slice with a thickness of a single cell size (0.535 Mpc) from our Galaxies source model of a simulation box of $L=75$ Mpc and $N=$140$^3$ at $z=8$. {\sc Second, third and fourth columns:} the signal as seen by SKA, HERA and LOFAR, respectively. {\sc First row:} the angular resolution limited signal. {\sc Second row:} instrumental thermal noise realizations. {\sc Last row:} the addition of the thermal noise to the angular resolution-limited signal.}
\label{fig:noise_maps}
\end{figure*}
\subsubsection{Foreground cleaning}
We next adjust the 21cm brightness temperature boxes by removing $k$ modes that are contaminated by foregrounds. These $k$ modes come in the form of a ``foreground wedge" in the $k_{\perp}$-$k_{\parallel}$ plane (see, e.g., \citealt{liu14} for more details on the wedge and its complement, the reionization window), where $k_{\parallel}$ is the comoving wavenumber in the direction parallel to the line of sight.  The wedge can be parameterized by saying that modes satisfying
\begin{equation}
k_\parallel  \leq m\, k_\perp,
\end{equation}
are foreground contaminated, where $m$ is a wedge slope that is given by
\begin{equation}\label{eq:wedge_eq}
m = \frac{ D \, H_{0}\, E(z)\, \sin \theta}{c(1+z)},
\end{equation}
where $H_{0}$ is the Hubble parameter, $c$ is the speed of light, $E(z)\equiv\sqrt{\Omega_{m}(1+z)^{3} +\Omega_\Lambda}$, and $\theta$ is the beam angle. A conservative way to fight foregrounds is simply to zero all modes within the wedge.

Using Equation \eqref{eq:wedge_eq}, the wedge slope $m$ value is about $m \sim 3$ during reionization redshifts, if one assumes that $\theta=90^{\circ}$ (horizon limit) for all experiments. However, we also consider smaller values of $m$ to model foreground mitigation algorithms that are able to reduce the Fourier space footprint of the contaminants. Such algorithms reduce the number of modes that need to be zeroed out, preserving more of the information content of the data.

To mimic the effect of foreground contaminants (and the effect of their removal) in our boxes, we proceed as follows:
\begin{itemize}
\item We compute the wedge slope $m$ at the appropriate redshift $z$ using Equation~\eqref{eq:wedge_eq} and additionally consider wedge slopes that are smaller than this value.
\item We zero out all Fourier modes satisfying $k_\parallel  < m\, k_\perp$.
\end{itemize} 
\begin{table*}\Huge
\hspace*{-1.5em}
 \scalebox{0.5}{\begin{tabular}{ l  c  c c }\hline
   {\bf Experiment } &  {\bf LOFAR} &  {\bf HERA} & {\bf SKA} \\ \hline \hline
    Antennae design & 48 tiles of bow-tie high band   & 350 hexagonally packed  & 866 compact core \\ \hline
    Antenna diameter, $D$ [m] & 30.75 & 14 & 35 \\ \hline
    Collecting area [m$^2$] & 35,762 & 50,953 & 833,189 \\ \hline
    system temperature [K] ($=T_{\rm sky} + T_{\rm rcvr}$) & T$_{\rm sky}$ + 140 & T$_{\rm sky}$ + 100 & 1.1 T$_{\rm sky}$ + 40  \\ \hline
    Dish area, $A_{\rm dish}\,\, [m^{2}] = \pi(D/2)^{2}$ & 742.64 & 153.93&  962.11 \\ \hline
    Integration time T$_{\rm int}$ [s] & 60 & 60 & 60  \\ \hline
    Band width (B) [MHz] & 8 & 8 & 8 \\ \hline
    	Number of channels ($N_{\rm chan}$) & 82 & 82 & 82 \\ \hline
    	frequency resolution $\Delta nu = {\rm B}/N_{\rm chan}$ [KHz]& 97.6 & 97.6 & 97.6 \\ \hline 
    	Number of observing hours/days & 6/180& 6/180 & 6/180 \\ \hline
    	Redshift/frequency [MHz] & 8 / 157.82 & 8 / 157.82 & 8 / 157.82 \\ \hline
    	Beam angle $ \theta \,\,[\rm rad]$    & 0.618   & 1.358  &  0.543   \\ \hline
    	Default wedge slope $m$, Equation.~\eqref{eq:wedge_eq} & 2.0 & 3.4 & 1.8 \\ \hline
    	Maximum $k_{\perp}$ [h Mpc$^{-1}$] & 2.0 & 0.6 & 3.6 \\ \hline
    Averaged thermal noise error, $\sqrt{\langle|N|^{2}\rangle}\, [\rm Jy]/[\rm mK]$ & 0.69/236   & 3.03/215  & 0.45/200\\ \hline
\end{tabular}}
\caption{Summary of parameters used in this study to obtain the thermal noise sensitivity for each experiment.}\label{array_tab}
\end{table*}

\begin{figure*}
\centering
\setlength{\epsfxsize}{0.5\textwidth}
\centerline{\includegraphics[scale=1.]{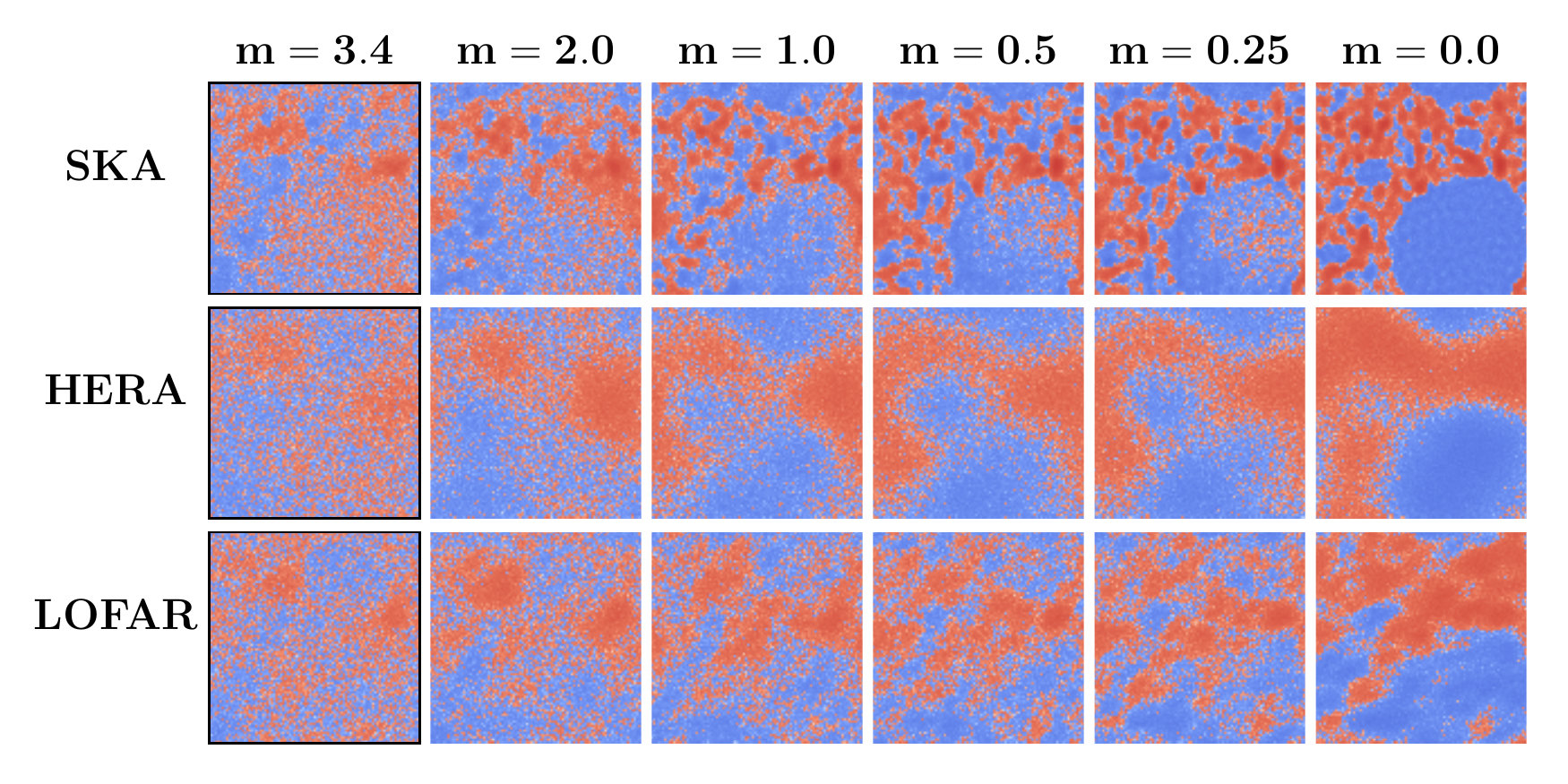}}
\caption{ Same 21cm signal as in Figure~\ref{fig:noise_maps}, after the angular resolution and thermal noise treatment, but with different wedge filtering levels, as quoted in subtitles. {\sc First, second and third rows} are the signal as seen by the SKA, HERA and LOFAR, respectively. At a wedge slope $m \lesssim  0.5$, the signal features start to appear more clearly, particularly for the SKA.}
\label{fig:wedge_maps}
\end{figure*} 

\subsubsection{Thermal noise}
 The thermal noise error of the observed visibilities in the $uv$-plane are uncorrelated between measurements~\citep{fur06}, and hence can be drawn from a Gaussian distribution of a zero mean and a standard deviation~\citep{zal04} given by: 

\begin{equation}\label{eq:sigma_noise}
\sqrt{\langle|n|^{2}\rangle} [{\rm Jy} ] =  \frac{ 2 \,k_{\rm B} \,T_{\rm sys} }{ A_{\rm dish} \sqrt{\Delta\nu \,T_{\rm int}}  }\,\, ,
\end{equation}
where $\langle|n|^{2}\rangle$ is the averaged noise squared, the Boltzmann constant is $k_{\rm B} = 1.38\times 10^{-23}\, [{\rm J/K}]$,  $A_{\rm dish}$ is the area of a single antenna, and $T_{\rm int}$ is the integration time to observe a single visibility at a frequency resolution $\Delta\nu$. The $T_{\rm sys}$ is the system temperature which is a combination of the sky temperature $T_{\rm sky}$ and the receiver temperature $T_{\rm rcvr}$. We adopt the commonly used power law model~\citep{thoms07} for $T_{\rm sky}$ as a function of frequency such that  $T_{sky} = 60( \nu / 300.0  )^{-2.55} \, [\rm K]$, which agrees with low frequency observations. Equation~\eqref{eq:sigma_noise} provides the flux error in unit of Jansky which we convert to Kelvin using:
\begin{equation}
\sqrt{\langle|n|^{2}\rangle} [\rm K] =  \frac{\lambda^{2}}{2\, k_{\rm B}\, \Omega}\times \sqrt{\langle|n|^{2}\rangle} [\rm Jy]  = \frac{ \lambda^{2} \,T_{\rm sys} }{ A_{\rm dish} \,\Omega\, \sqrt{\Delta\nu \,T_{\rm int}}  } \, ,
\end{equation}
where $\Omega \approx \theta^{2}$ is the beam area and $\lambda$ is the signal wavelength at the observation redshift $z$.  The receiver temperature, with other parameters, is listed in Table~\ref{array_tab}. 

Our recipe to generate the 2D thermal noise realization is as follows:
\begin{itemize}
\item We first generate 2D grid (in ${\bf k}_{\perp}$ direction) from a Gaussian distribution of a mean zero and a standard deviation $\sqrt{\langle|N|^{2}\rangle}$ both independently in the imaginary and real parts in Fourier space.
\item At non-zero $uv$ coverage pixels, we suppress the noise by the amount of the $uv$ coverage (baselines intensity) as $1/\sqrt{N_{uv}}$.
\item We reduce the noise further over the total observation days by a factor of  $1/\sqrt{N_{\rm days}}$.
\item We inverse Fourier transform the thermal noise to the real space.
\end{itemize}

We note that a similar pipeline has been previously developed in~\citet{ghara17}.

As a final step, we add the noise realization  to our cosmological signal image (which has been treated for angular resolution and foreground effects, as discussed above). This then forms our foreground-removed 21 cm observations

Figure~\ref{fig:noise_maps} illustrates the impact of instrumental effects. There, we show a slice from a box based on our Galaxies model at $z=8$ (bottom left), the same slice but angular resolution limited (excluding $k$-modes at zero $uv$ coverage; first row), the thermal noise contributions from each experiment (second row), and finally the image as observed by SKA, HERA, and LOFAR, respectively. In the first row, we see the effect of angular resolution, which is set by the detailed baseline distribution in the $uv$-plane. As seen in Figure~\ref{fig:uv_map}, the SKA has the highest $uv$ coverage that extend down to very small angular scales of about 3.6 h Mpc$^{-1}$, and hence its angularly limited version (top of second column) is quite similar to the original signal (bottom left). The compact design of HERA also obtains a high $uv$ coverage but on somewhat larger angular scales up to 0.6 h Mpc$^{-1}$, which prevents the feature resolution on smaller scales  of the signal, but bubbles can still be detected on large scales  (top of third column). While the $uv$ coverage of LOFAR extends to smaller scales down to 2 h Mpc$^{-1}$, the $uv$ coverage is low due to the sparse baseline distribution. This leads to a poorer angular resolution on both large and small scales features (top of forth row). Next, we produce thermal noise maps (second row). As expected, the SKA has a slightly dimmer noise map than HERA and LOFAR do. Finally, we add the noise maps to the angular resolution-limited to produce mock images as seen by our experiments (last row). We see that angular resolution has a higher impact on the signal than the thermal noise does, since the angular resolution limited signal images are similar to the final images after adding the thermal noise, albeit HERA and LOFAR are more noisy than SKA due to the higher thermal noise.
Because of the low angular resolution, we expect that LOFAR may not be able to easily distinguish between images generated from the Galaxies and AGN scenarios.
 
In Figure~\ref{fig:wedge_maps}, we explore the effect of different wedge filtering levels. The default wedge slope $m$ values, from equation~\eqref{eq:wedge_eq}, are 2.0, 3.4 and 1.8 for LOFAR, HERA and SKA respectively. Since these values are quite large, we consider rather a range of wedge slopes to gradually display the effect of foreground contamination level based on our method. Hence, we show the signal's response to different wedge slopes $m=3.4$, 2.0, 1.0, 0.5, 0.25, 0.0 (no wedge applied, 100\% successful foreground removal). Results for the SKA, HERA and LOFAR are shown in the first, second, and third row, respectively. As expected, the lower the wedge slope, the clearer the signal features are, since fewer $k$-modes are removed due to foregrounds. At $m < 0.5$, the signal prominent features on large scales begin to appear more clearly, and hence we expect better classification performance at this limit of wedge filtering, particularly for the SKA. On the other hand, adding any amount of foreground contamination, even as very low as using $m=0.25$, destroys the signal features, on large spatial scales along $k_\parallel$ and on small spatial scales along $k_\perp$, for HERA and LOFAR, and hence we expect that these arrays might only be able distinguish between our Galaxies and AGN models at $m=0.0$, which corresponds to 100\% successful foreground removal. In Section \ref{sec:results}, we will explore the classifier ability to discriminate between Galaxies and AGN models as a function of the wedge slope.

\section{Reionization Models Classifier Architecture}\label{sec:classifier}

\begin{figure*}
\centering
\setlength{\epsfxsize}{0.5\textwidth}
\centerline{\includegraphics[scale=0.55]{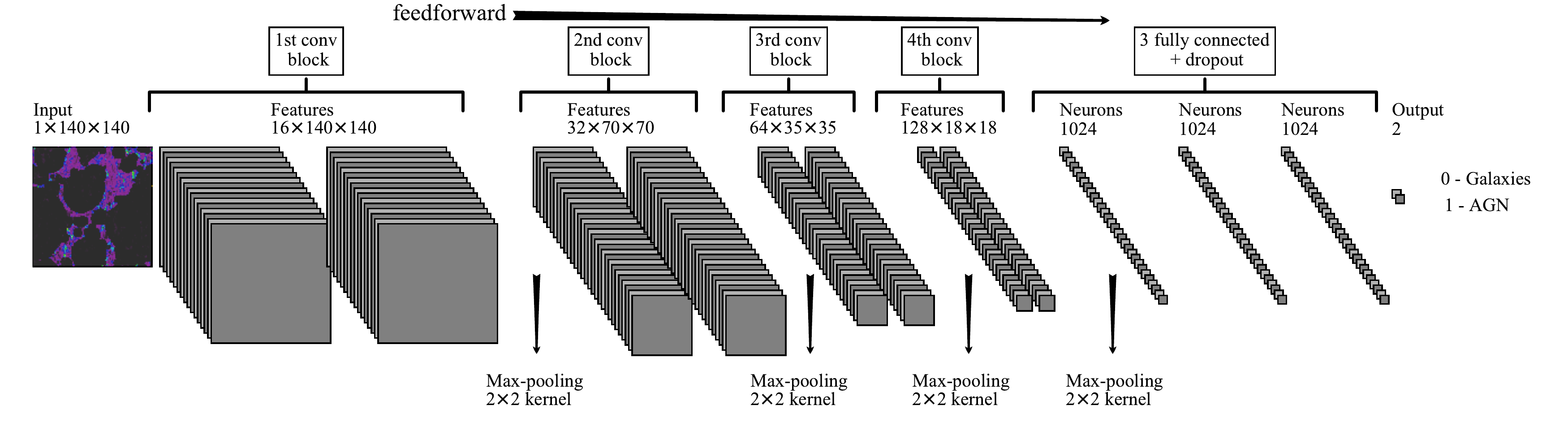}}
\caption{Reionization models classifier architecture. Each 21cm map is processed into four blocks of two convolutional and pooling layers that are followed by three fully connected layers to eventually output the image class (0-Galaxies, 1-AGN). Dropout regularizations are only applied on the fully connected layers during training, keeping 75\% of the neurons. Batch normalization is applied before all layers except the final output layer.}
\label{fig:cnn_arch}
\end{figure*} 

We employ Convolutional Neural Networks (CNNs) to build up the reionization models classifier. CNNs are a class of deep learning models that are powerful for large scale image recognition~\citep[for a comprehensive review see][]{rawang}. Similar to the classical Artificial Neural Networks (ANNs), CNNs are feedforward networks that take in an input ($\mathbf{x}$) and eventually outputs a predicted label ($y$). The connection between these is constructed by stacking neuron layers, each of which is a linear combination of the components of $\mathbf{x}$ with weights and biases. Each layer can produce multiple outputs of the form
\begin{equation}\label{eq:ann}
y^{j} = \phi \left(\sum^{N}_{i} w_{i}^{j}\, x_{i} + b^{j}\right),
\end{equation}
where $N$ is the number of inputs in the layer, $y^j$ is the $j$th output (or ``feature"), $w_i^j$ is the weighting of the $i$th input in the $j$th feature, and $b^j$ is the bias in the $j$th feature. The function $\phi$ is a nonlinear function response function that is chosen by the architect of the neural network.
\begin{table}\centering \LARGE
 \scalebox{0.5}{\begin{tabular}{ c  c  c }\hline
   {\bf Step } & {\bf Layer type} & {\bf Output dimension} \\ \hline \hline
     1 &  3$\times$3 Convolution  & 16$\times$140$\times$140 \\ \hline 
     2 & Batch Normalization + ReLU &   $-$ \\ \hline
    3  & 3$\times$3 Convolution   & 16$\times$140$\times$140  \\ \hline 
     4 & Batch Normalization + ReLU & $-$  \\ \hline 
     5 & 2$\times$2 max-pooling  & 16$\times$70$\times$70 \\ \hline
     6 & 3$\times$3 Convolution  & 32$\times$70$\times$70 \\ \hline
     7 & Batch Normalization + ReLU  & $-$ \\ \hline 
     8 & 3$\times$3 Convolution    & 32$\times$70$\times$70 \\ \hline
     9 & Batch Normalization + ReLU & $-$  \\ \hline 
     10 & 2$\times$2 max-pooling  & 32$\times$35$\times$35 \\ \hline
     11 & 3$\times$3 Convolution   & 64$\times$35$\times$35 \\ \hline
     12 & Batch Normalization + ReLU  & $-$ \\ \hline 
     13 & 3$\times$3 Convolution    & 64$\times$35$\times$35 \\ \hline
     14 & Batch Normalization + ReLU   & $-$  \\ \hline 
     15 & 2$\times$2 max-pooling  & 64$\times$18$\times$18 \\ \hline
     16 & 3$\times$3 Convolution  & 128$\times$18$\times$18 \\ \hline
     17 & Batch Normalization + ReLU & $-$  \\ \hline 
     18 & 3$\times$3 Convolution   & 128$\times$18$\times$18 \\ \hline
     19 & Batch Normalization + ReLU  & $-$ \\ \hline 
     20 & 2$\times$2 max-pooling  & 128$\times$9$\times$9 \\ \hline
     21 & Flattening  & 10368 \\ \hline
     22 & Fully connected   & 1024 \\ \hline
     23 & Batch Normalization+ ReLU    & $-$  \\ \hline 
     24 & Dropout (25\%)  & $-$  \\ \hline
     25 & Fully connected  & 1024 \\ \hline
     26 & Batch Normalization+ ReLU    & $-$  \\ \hline
     27 & Dropout (25\%)  & $-$  \\ \hline
     28 & Fully connected & 1024 \\ \hline
     29 & Batch Normalization + ReLU  & $-$  \\ \hline 
     30 & Dropout (25\%)   & $-$ \\ \hline
     31 & Fully connected + softmax & 2 \\ \hline
\end{tabular}}
\caption{Architectural summary of our classifier that is used to map the input 140$\times$140 21cm images into binary model classes (0-Galaxies, 1-AGN).}\label{cnn_step}
\end{table}
\begin{figure*}
\centering
\setlength{\epsfxsize}{0.5\textwidth}
\centerline{\includegraphics[scale=1.5]{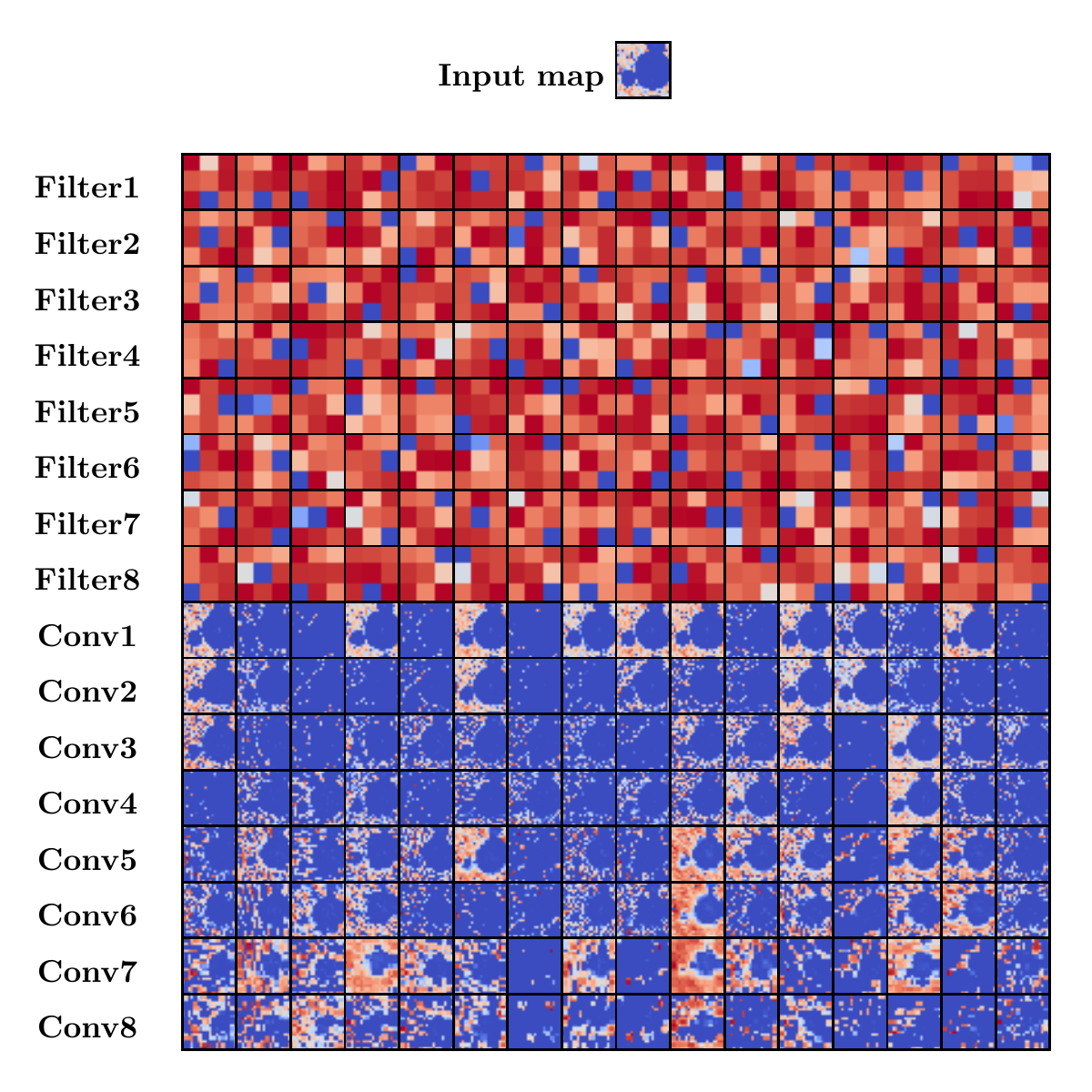}}
\caption{The evolution of a single 21cm map ({\sc top-middle}) through the first 16 filters of the classifier layers as stored in the final step of training, for the SKA model in Section \ref{sec:2nd_case}. Titles indicate the filters and the results of the convolutions (conv), including the application of the ReLU activation function, which sets all negative pixels to zero as shown by the blue color. The numbers indicate the order of the convolutional layers in the network as shown in Figure~\ref{fig:cnn_arch}. Some filters detect small scale features but most of them strongly highlight the large scale bubbles. It is evident that the classification is driven by the large-scale bubbles.}
\label{fig:processed}
\end{figure*}

The CNN architecture is comprised of a series of different type of neuron layers that each have a specific form for the weights:
\begin{itemize}
\item {\bf Convolutional layers} extract different features by convolving the input image with learned weights of 2-dimensional kernels.
\item {\bf Pooling layers} reduce the dimension and spatial resolution of features. They speed up the network performance and prevent over-fitting. Examples of pooling layers are average-pooling or max-pooling in which the input image is divided into non-overlapping sub-images where the values of each sub-image are the average or max of its original values from the input image, respectively.  
\item {\bf Fully connected layers} are the classical ANN layers in which the input is connected to all neurons.
\end{itemize}
Two other standard techniques are employed in our CNN:
\begin{itemize}
\item {\bf Dropout} regularizes the training process by randomly switching off neurons with a given probability.
This prevents coadaptations of same features and overcomes the problem of overfitting.
\item {\bf Batch Normalization} insures that the input images, before each layer, are uncorrelated while keeping the averages $\mu$ and variances $\sigma$ fixed. Batch normalization scales and shifts the input by two new learned parameters $\gamma$ and $\beta$ as follows:
\begin{equation}
y_{i} = \gamma \frac{x_{i} -\mu}{\sqrt{\sigma^{2} + \epsilon}} \beta,
\end{equation}
where $\epsilon$ is a small constant to avoid arithmetical instability. This technique has shown to improve training performance significantly~\citep[e.g.][]{roy18,sch18}.
\end{itemize}
Combining this together, our CNN architecture is as follows:
the network is composed of four blocks of convolutional layers, three fully connected layers, and an output layer. Similar CNN architectures have been very successful to perform large-scale image recognition~\citep[e.g.][]{simon15,sch18}. In each block, there are two convolutional layers with convolutional kernels of size 3$\times$3 and a single 2$\times$2 max-pooling layer. The number of generated features increases from one block to another while the features' dimension decreases through the max-pooling. The final output layer with two neurons is applied to classify the input 21cm map by providing the probabilities that the input images were generated from the Galaxies or AGN models. Dropout is applied on all fully connected layers (only during training) to keep 75\% of the neurons. Before each layer, batch normalization is applied except for the final output layer where a softmax activation is applied instead. 
The softmax function reads as follows:
\begin{equation}
S(x^{i}) = \frac{\exp{x^{i}}}{\sum_{j}\exp{x^{j}}}\,.
\end{equation}

 The Rectified Linear Unit (ReLU) activation function is applied on all convolutional and fully connected layers. The ReLU function is defined as:
\begin{equation}
\rm ReLU(x) = max(x,0),
\end{equation} 
which returns $0$ if the input $\rm x$ is negative; otherwise it keeps the raw input.
 Figure~\ref{fig:cnn_arch} shows our classifier design and Table~\ref{cnn_step} summarizes the image processing steps through the classifier layers as outlined above.

At the beginning of training, we need to initialize all layers weights and biases with different small numbers to break the symmetry between layers neurons. There are different initialization strategies, and the most common approach is the use of random numbers sampled from a Gaussian distribution of zero mean and finite unitless  standard deviation (e.g., $\sigma$ = 0.1 or 0.01 ). However, in order to ensure faster training and keep the input data properties, particularly, the variance, we initialize all weights and biases with a generalized form of Xavier initializer~\citep{xav10} that is called the Variance Scaling initializer. In such an initialization, the random numbers are drawn from a zero mean Gaussian distribution with a variance equal to the inverse of the average of the number of input and output neurons. This initializer insures that the variance of the input data is preserved through all the classifier layers.

At each training step, we first calculate the distance between the true and predicted labels.  While the simple Cartesian distance ($\sqrt{x_{\rm true}^{2} - x_{\rm predicted}^{2} }$) often works, many studies found that the cross entropy ( $- x_{\rm true} \log(x_{\rm predicted})$) is more numerically stable measure, and hence we use the cross entropy as our loss function. We next minimize the loss function using a mini-batch gradient descent momentum optimizer. With this optimizer, the loss gradients are computed with respect to the classifiers' parameters (weights, biases and the batch normalization parameters) for a mini-batch of the training dataset to indicate in which directions of parameter space one should move in order to approach the global minimum. The update is controlled by the learning rate, which speeds/slows the learning process, and with the momentum term, which accelerates the update direction by keeping a fraction of the previous update direction. This fraction is called momentum and is usually set to a value close to unity. At zero momentum, the Momentum optimizer becomes the standard Gradient Decent optimizer. Here, we use a value of 0.9 for the  unitless momentum and 0.01 for the  unitless  initial learning rate, which then exponentially decays by 97\% for every 20 training steps. 

It is worthwhile to mention that the architecture in this paper is different from and more complex than previous architectures used to classify the same models in~\citet{hassan_iau}. In~\citet{hassan_iau}, the training datasets were obtained by varying the models' astrophysical parameters while leaving the underlying density realization fixed. The treatment of instrumental effects was also considerably simpler. As a result, a very simple network architecture, with two convolution blocks and a single fully connected layer, could produce high accuracy ( > 95\%) from the first few training epochs. With the more complicated training sets of this paper, deeper and more complex networks are necessary.
 
To illustrate how the classifier extracts different features from the input 21cm images, we show in Figure~\ref{fig:processed} an example of a single 21cm image as it passes through the different layers of the trained neural network. We show 
only the first 16 filters (i.e., convolution kernel) from each convolutional layer. We also show the convolution
of each filter with the input image, together with the application of the ReLU activation function. We see that different filters extract subtly different features of the ionised bubbles from the input image. The input image responds 
differently to each filter and the convolution output shows different levels of bubbles detections. The ReLU
activation function then clears away the less important features (negative pixels) and keeps the most prominent
features (positive pixels). This process continues through all convolutional layers. The eighth convolutional layer 
is the last step before the two-dimensional information is collapsed into an abstract data vector. At this layer (Conv8; last row), we see that most of the filters strongly highlight the large scale bubbles while a few of the filters extract small scale features. In addition, the max-pooling layer is applied three times before the last convolutional layer, which reduces the input image dimension from 140$\times$140 to 18$\times$18, and hence smooths the small scale features more than the large scales ones. This indicates that the classification might be primarily determined by the sizes of large-scale bubbles. As we saw in Figure~\ref{fig:wedge_maps}, the prominence of large-scale bubbles depends on the effectiveness of foreground removal.  For the SKA, a foreground cut higher than $m>0.5$ makes ionised bubbles difficult to identify, and the large scale features reappear at lower $m$ values. Any amount of added foreground contaminants washes away the remaining signal features in the case of HERA and LOFAR, since the angular resolution filtering of the signal already removed most of the signal prominent features.  At 100\% successful foreground cleaning ($m=0.0$), the signal features can be easily recognized even in the presence of thermal noise levels representative of our experiments.

\section{Training dataset}\label{sec:dataset}
We generate our training dataset from a simulation box of size 75 Mpc with 140$^{3}$ cells. We choose this box size since it  partially covers the relevant scales that future 21cm surveys will be sensitive to. In addition, we have previously tested the 21cm power spectrum convergence with the box size and found that our simulations are well converged down to a box size of 75 Mpc~\citep{hassan16}.

We run 1,000 reionization simulation realizations with 1,000 different random seed numbers for the initial density field fluctuations from each model. We choose 1,000 values out of the following parameters ranges:
\begin{itemize}
\item {\bf Galaxies}: (0.5$\geq \fescs \geq$0.01), (10$^{40}$ $\geq A_{\rm \star} \geq$ 10$^{39}$), (1.0$\geq C_{\rm \star} \geq$0.0).

\item {\bf AGN}: (1.0$\geq \fesca \geq$0.5), (10$^{7}$ $\geq A_{\rm AGN} \geq$ 10$^{5}$), (1.0$\geq C_{\rm AGN} \geq$0.0).
\end{itemize}
These parameters range translate into a spread in the total ionizing emissivity of each source model. For instance with this range, at $z=5$, Galaxies have a total ionization rate range of $\log(\dot{N}_{\rm \star,tot}) =  (- 2.5, 2.7)  \,[\rm photons/seconds/Mpc^{3}]$, whereas the AGN total ionization rate has a range from $\log(\dot{N}_{\rm AGN, tot}) =  (-1.5, 8.9) \,[\rm photons/seconds/Mpc^{-3}]$. These ranges are much larger than the $1\sigma$ level of the ionising emissivity measurements at $z=5$ by~\citet{bec13}, which is $\log(\dot{N}_\textrm{tot}) = -0.014^{+0.454}_{-0.355}\,[\rm photons/seconds/Mpc^{-3}]$. By neglecting such additional constraints in this paper, we are conservatively considering a wider parameter space than is currently allowed by observations. In future work, one hopes that the job of a classifier can be made easier by using ionising emissivity observations to restrict the possible parameter space of models.

In addition to the variations in the astrophysical parameters, our training datasets also have varied underlying density fields (by initializing each box with a different random seed). Our training set is thus much more complex than the ones previously considered in~\citet{hassan_iau} and more suitable for training classifiers that might be applied to realistic 21cm surveys.
\begin{figure*}
\centering
\setlength{\epsfxsize}{0.5\textwidth}
\centerline{\includegraphics[scale=0.6]{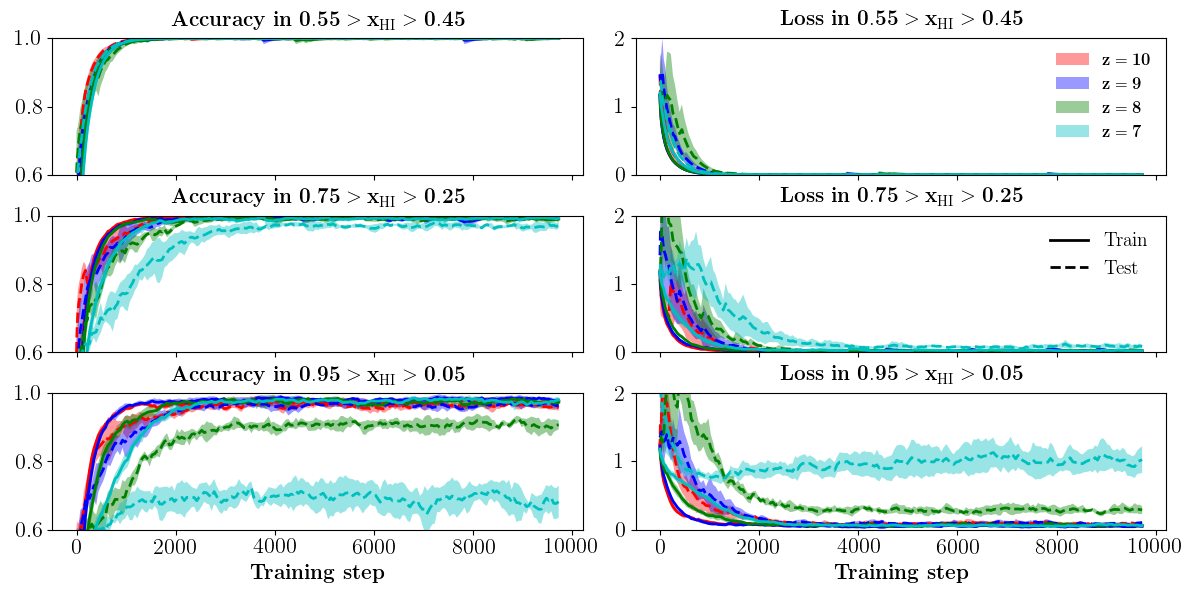}}
\caption{Classifier accuracy and loss as a function of training step for different redshifts and different neutral fraction range. Solid and dashed lines are the mean accuracy of training and testing samples. Red, blue, green, cyan shaded areas are the 1$\sigma$ level out of the 5 runs with different random number generator to initialize the classifier weights. The classifier converges after the first $\sim2000$ training steps ($\sim$ 5 training epochs). Accuracy is greatest at highest redshifts, or when datasets are restricted to narrow neutral fraction ranges. Except at $z=7$ (when the universe is highly ionized), the classifier is able to discriminate between models at an accuracy of about 92-100\%, depending on redshift and neutral fraction range.}
\label{fig:zvsxHI}
\end{figure*} 
\begin{table*}\LARGE
\hspace*{-1.5em}
\resizebox{\textwidth}{!}{\large
 \begin{tabular}{ | c|   c|  c | c | c|}\hline
   & \multicolumn{4}{|c|}{\bf Accuracy}  \\ \hline \hline
  {\bf $x_{\rm HI}$ range} & {\bf z=7} &{\bf z=8} &  {\bf z=9}& {\bf z=10} \\ \hline \hline
  {\bf 0.55 > x$_{\rm HI}$> 0.45 } &  1.0000 $\pm$ 0.0000   &  0.9971 $\pm$ 0.0018 & 1.0000 $\pm$ 0.0000  & 1.0000 $\pm$ 0.0000 \\ \hline  
   
   {\bf 0.75 > x$_{\rm HI}$> 0.25 } & 0.9680 $\pm$ 0.0031 & 0.9875 $\pm$ 0.0020 & 0.9925 $\pm$ 0.0021 & 0.9969 $\pm$ 0.0012 \\ \hline 
   
 {\bf  0.95 > x$_{\rm HI}$> 0.05} & 0.6662 $\pm$ 0.0109 & 0.9159 $\pm$ 0.0096 &  0.9608 $\pm$ 0.0025 & 0.9642 $\pm$ 0.0018 \\ \hline 
\end{tabular}}%
\caption{Summary of the classifier accuracy and loss as a function of redshift and neutral fraction at the final training step as evaluated on the validation dataset. Results are quoted in terms of the mean and standard devation of the five classifier runs with different random number generator to initialize the weights. Instrumental effects are not taken into account.}\label{case1}
\end{table*}
The specific ranges chosen for the Galaxies parameters come from~\citet{hassan17}. The ranges are essentially the 1$\sigma$ level obtained in an MCMC parameter estimation to match various reionization observables. As for the AGN source model, the AGN photon escape fraction $\fesca$ is most commonly assumed to be unity due to the observed hardness of AGN spectra. However, we relax this assumption and consider as lower values as 50\%. In~\citet{hassan18}, we showed that the AGN model reproduces ionising emissivity constraints if $ A_{\rm AGN} \sim$ 10$^{6}$, and hence we consider one order of magnitude above and below this value. The emissivity's non-linear dependence on black hole mass $C_{\rm AGN}$ follows our choice of the similar parameter $C_{\rm \star}$ from Galaxies models. However, as previously shown in~\citet{hassan17}, these parameters are degenerate, particularly between the photon escape fraction and emissivity amplitude, so simulation realizations with values lie outside these ranges can easily be extrapolated. 

We note that, from each simulation, one may extract 140 images of 21 cm fluctuations, choosing one of the principal axes (x, y, or z) of the simulation as the line-of-sight axis. A two-dimensional image is generated by extracting a slice with the line-of-sight axis having a thickness of a single cell (0.535 Mpc), and the full extent of the box (75 Mpc) along the other two axes. This shows that 1,000 simulation realizations, in fact, contain 140$\times$3$\times$1000 $=$ 420,000 possible different 21cm images, which is sufficient for training a deep network. However, close 21cm slices from the same box might share similar features, and hence we jump at least for $\sim$ 2 Mpc comoving distance\footnote{ This is the minimum distance in our simulations beyond which slices are not correlated, which we choose in order to support our training samples with as distinct features as possible.}  before we extract the following slice. Out of these 1,000 boxes, there are around $\sim$ 300 boxes that produce a neutral fraction between 95\% $> x_{\rm HI} >$ 5\% at different redshifts, and hence we limit our study to those boxes since we expect our two models to produce very similar 21cm fluctuations at the beginning (x$_{\rm HI}$ > 95\%) and end (x$_{\rm HI}$ < 5\%)  of reionization. 

After generating our training samples, we use $\sim$ 8\% for testing, $\sim$8\% for validation, and $\sim$86\% for training. For each study, we use at least 6,000 different images of each class for training and at least 1,000 different images of each class for testing and validation.  At each training step, we randomly select a batch of 32 labelled images for training and testing. We have tested the effect of selecting small versus large batches and the result remains unchanged.

\section{Results}\label{sec:results}

We quantify the classifier performance by its accuracy, which is defined as the percentage of images that are recognized correctly. For each section, we run the classifier five times using different random seeds for initializing the learned parameters of the network. This examines the classifier ability to converge to same results irrespective of the random initializations of weights. For these five different runs, we use the validation sample to evaluate the classifier accuracy at the end of training. We then express all results in terms of the mean accuracy and standard deviation out of these five runs.

\subsection{Dependence on redshift and neutral fraction (without noise and foreground removal)}\label{sec:1st_case}
We first study the classifier accuracy and loss ( the cross entropy as described earlier which is similar to the error rate) as a function of redshift and neutral fraction,  without including noise and foreground effects. The results are shown in Figure~\ref{fig:zvsxHI}. The red, blue, green, and cyan colours represent the redshifts 10, 9, 8, and 7, respectively. Solid lines indicate the mean accuracy/loss from training samples whereas the dashed lines are from testing samples. Shaded areas show the 1$\sigma$ regions out of the five runs with different random seeds. In all cases, the classifier converges within the first 2000 training steps ($\sim$ 5 training epochs) to yield accuracies and losses that stay approximately constant. The 1$\sigma$ levels (shaded areas) are quite small when one is restricted to narrow neutral fraction ranges. 
\begin{table*}\Huge
\hspace*{-1.5em}
 \scalebox{0.45}{\begin{tabular}{ | c | c | c | c| c | c |}\hline
   {\bf Test }    & \multicolumn{5}{|c|}{\bf Accuracy}   \\ \hline \hline
   {\bf No Instrumental Effects} & \multicolumn{5}{|c|}{\bf 0.990 $\pm$ 0.002} \\ \hline 
    & & \multicolumn{4}{|c|}{\bf Angular resolution + Thermal noise } \\ \hline
    	& {\bf Angular resolution} & {\bf m=0.0} & {\bf m=0.25} & {\bf m=0.5} & {\bf m=1.0}\\ \hline 
   {\bf SKA} &{\bf 0.984 $\pm$  0.001} & {\bf 0.976 $\pm$ 0.001 }& {\bf 0.977 $\pm$ 0.001} & {\bf 0.957 $\pm$ 0.004} &  0.860 $\pm$ 0.011\\ \hline
   {\bf HERA} &{\bf 0.952 $\pm$ 0.001} &{\bf  0.946 $\pm$ 0.004 }& 0.881 $\pm$ 0.005 & 0.851 $\pm$ 0.017 & 0.749 $\pm$ 0.005 \\ \hline
   {\bf LOFAR} &{\bf  0.920 $\pm$  0.007 }& {\bf 0.895 $\pm$ 0.001} & 0.839 $\pm$0.004 &  0.775 $\pm$ 0.017 & 0.672 $\pm$ 0.009\\ \hline 
\end{tabular}}
\caption{Summary of the classifier accuracy for different 21cm experiments at $z=8$ and neutral fraction range of  0.25 < x$_{\rm HI}$< 0.75 as a function of the wedge slope. Results are expressed in terms of the mean and standard deviation of the accuracy evaluated on the validation dataset at the final training step for the different five runs.}\label{case2}
\end{table*}
This shows that our classifier is robust and insensitive to the random initialization of network weights. We note that the classifier accuracy increases as the neutral fraction range decreases and redshift increases. We summarize the accuracy on the whole validation set for the 5 different runs in Table~\ref{case1}. When the universe is highly ionized at $z=7$, AGN-only and galaxies-only models tend to produce similar ionization and 21cm fields, and hence the accuracy is low ($\sim$66\%) on the testing and validation samples. In this case, we see that the loss increases slowly, which is an indication of slight over-fitting---with a large degree of similarity between these models at end of reionization, the network is resorting to simply memorizing the inputs and outputs of the training sets. Such an approach naturally fails for the test samples, so the loss rises. Excluding $z=7$ (when our Universe is highly ionized), the classifier is able to attain an accuracy of 92-100\% from the testing and validation samples across a wide variety of neutral fractions.

\begin{figure}
\centering
\setlength{\epsfxsize}{0.5\textwidth}
\centerline{\includegraphics[scale=0.6]{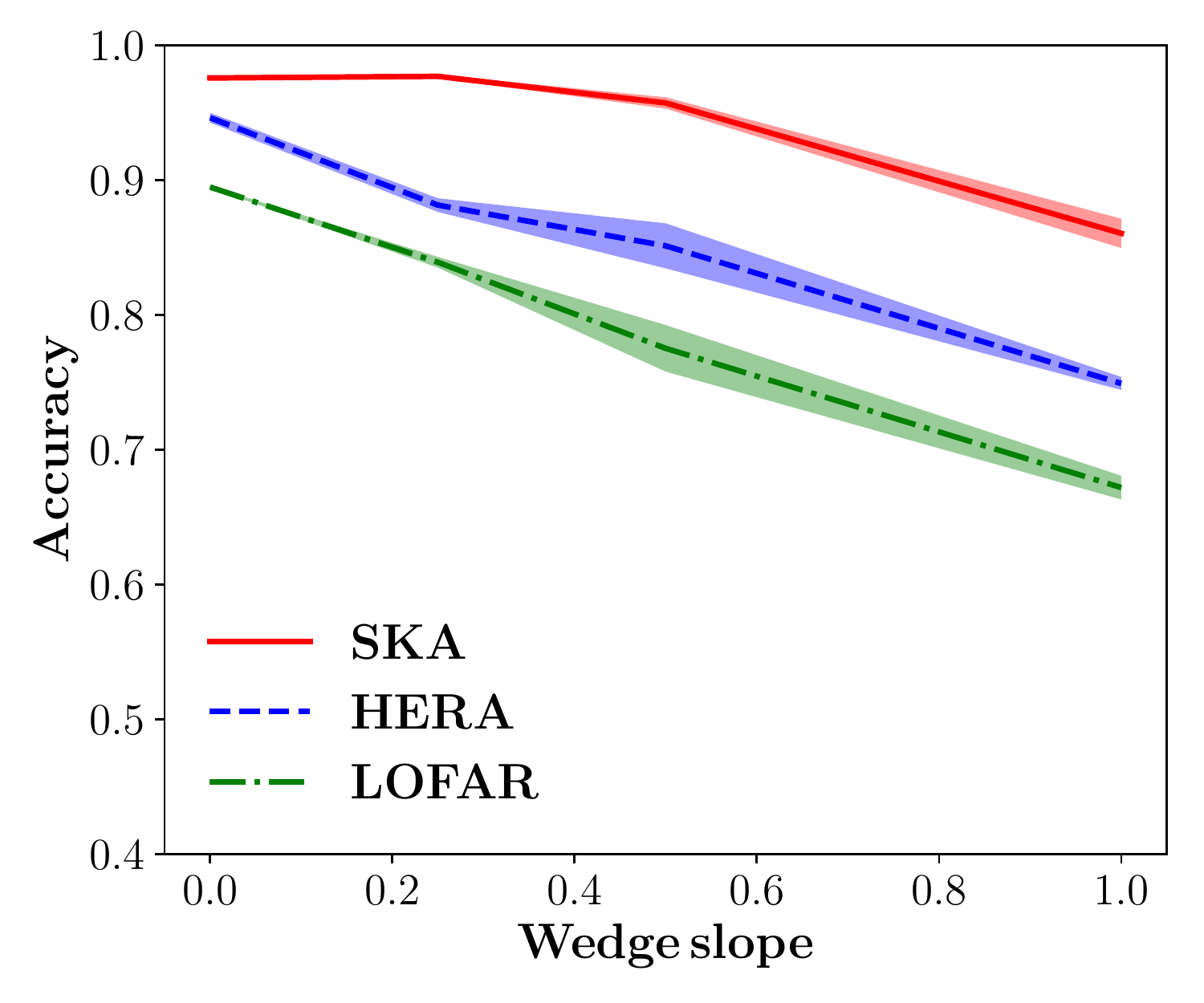}}
\caption{ Classifier accuracy as a function of the wedge slope for different 21cm experiments at fixed redshift $z=8$. The red solid, blue dashed, and green dash-dotted lines are the mean validation accuracy from the mock datasets as observed by the SKA, HERA, and LOFAR. Shaded areas are the 1$\sigma$ level of the validation accuracy for the different five runs. It is evident that 100\% successful foreground removal (m=0.0) is required for these experiments to achieve the desired accuracy, except the SKA can still discriminate between source models at stronger foreground contamination levels $m < 1.0$ .}
\label{fig:acc_noise}
\end{figure}
 
\subsection{Classification at a fixed redshift (with noise and foreground removal)}\label{sec:impace_noise}

We now determine the classifier accuracy in the presence of instrumental effects. To illustrate the impact of these effects, we use the dataset from $z=8$, restricted to the neutral fraction range of  0.75 > x$_{\rm HI}$> 0.25.  We then add noise and instrumental effects from SKA, HERA, and LOFAR. For each experiment, we consider five datasets; the first is with the effects of the angular resolution only, and the rest includes the addition of the thermal noise with different wedge slopes $m = 0.0$ (100\% successful foreground removal), 0.25, 0.5, 1.0 to remove foreground-contaminated $k$-modes. We train the classifier on each of the five datasets five times with different random seeds for the generation of initial weights and biases.

We show the classifier accuracy  as a function of the wedge slope in Figure~\ref{fig:acc_noise}. The red solid, blue dashed and green dash-dotted lines are the mean validation accuracy on datasets as observed by SKA, HERA, and LOFAR respectively. The corresponding shaded areas are the 1$\sigma$ levels from five different runs. We see that the accuracy gradually increases as the wedge slope decreases as the signal features start to reappear more prominently. As quoted in Table~\ref{case2}, only accounting for the angular resolution effects reduces the accuracy, as compared to the case without adding instrumental effects (=0.99\%), by $\sim$ 1\%, 4\% and 7\% for the SKA, HERA and LOFAR, respectively. Nevertheless, all experiments are here able to score an accuracy of $\gtrsim 90\%$. However, the SKA, HERA, LOFAR are still able to score an accuracy of $\sim$ 98\%, 95\% and 90\% respectively, in the presence of their thermal noise on top of the angular resolution effects, but assuming that 100\% successful foreground removal is perfectly achieved using $m=0.0$.

As expected, adding more foreground contaminants with higher values of $m$ reduces the accuracy significantly, particularly for HERA and LOFAR. As mentioned before, the default wedge slope values for these experiments are quite large (see Table~\ref{array_tab}) at which our classifier would clearly fail to distinguish between the models. This is not surprising given the feature disappearance and the noisy images of these experiments at high $m$ values as depicted in Figure~\ref{fig:wedge_maps}.  Nonetheless, the SKA is still able to score an accuracy of more that 90\% for wedge slope $m < 1.0$. As seen from Figure~\ref{fig:wedge_maps}, a wedge slope value of $m\lesssim1.0$ is roughly when large-scale bubbles begin to be recognizable, at least in the case of the SKA. That the classifier achieves high accuracy ($> 90\%$) at this point is suggestive of the interpretation that the classifier determines the differences between these models based on large-scale bubble. We summarize the validation accuracy for the five different runs in Table~\ref{case2}.

 \begin{table*}\Huge
\hspace*{-1.5em}
 \scalebox{0.45}{\begin{tabular}{ | c | c | c | c| c | c |}\hline
   {\bf Test }    & \multicolumn{5}{|c|}{\bf Accuracy}   \\ \hline \hline
   {\bf No Instrumental Effects} & \multicolumn{5}{|c|}{\bf 0.999 $\pm$ 0.0001} \\ \hline 
    & & \multicolumn{4}{|c|}{\bf Angular resolution + Thermal noise } \\ \hline
    	& {\bf Angular resolution} & {\bf m=0.0} & {\bf m=0.25} & {\bf m=0.5} & {\bf m=1.0}\\ \hline 
   {\bf SKA} & {\bf 0.910 $\pm$ 0.004 }& {\bf 0.908 $\pm$ 0.002 }& 0.720 $\pm$ 0.024& 0.721 $\pm$ 0.015 & 0.666 $\pm$ 0.004\\ \hline
   {\bf HERA} &0.703 $\pm$ 0.005 & 0.695 $\pm$ 0.007&  0.578 $\pm$ 0.003 & 0.578 $\pm$ 0.012& 0.540 $\pm$ 0.002 \\ \hline
   {\bf LOFAR} & 0.664 $\pm$ 0.005 & 0.644 $\pm$ 0.009 & 0.576 $\pm$ 0.019&   0.516 $\pm$ 0.013 & 0.503 $\pm$ 0.006\\ \hline 
\end{tabular}}
\caption{Summary of the classifier accuracy for different 21cm experiments at a fixed power spectrum as a function of the wedge slope. Results are expressed in terms of the mean and standard deviation of the accuracy evaluated on the validation dataset at the final training step for five different runs.}\label{case3}
\end{table*}

\subsection{Classification at a fixed power spectrum (with noise and foreground removal)}\label{sec:2nd_case}

\begin{figure}
\centering
\setlength{\epsfxsize}{0.5\textwidth}
\centerline{\includegraphics[scale=0.6]{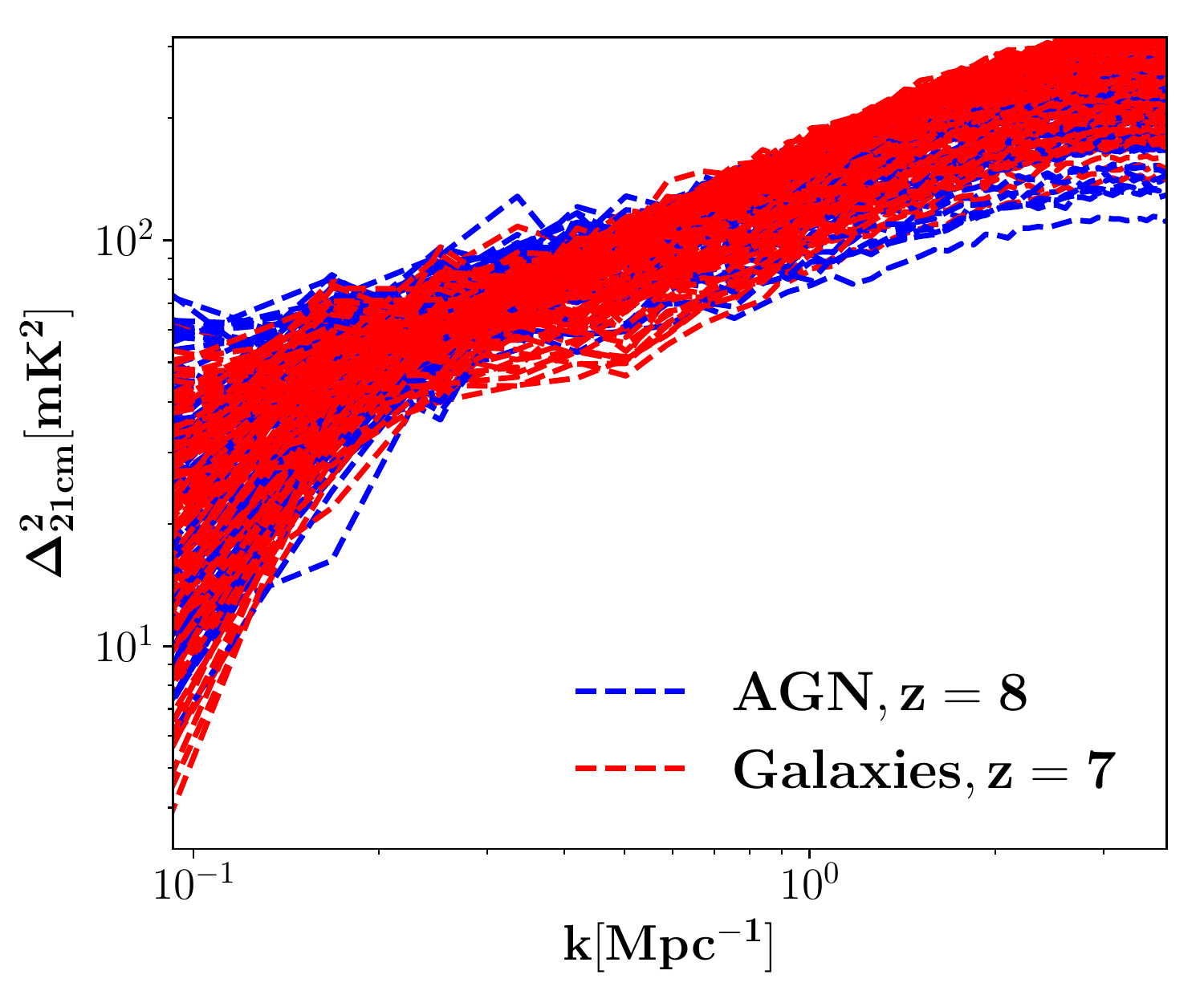}}
\caption{21cm power spectra from our Galaxies model at $z=7$ and our AGN model at $z=8$ in the neutral fraction range 0.45< x$_{\rm HI}$ < 0.95. At these redshifts and neutral fractions, the two models have similar 21cm power spectra and the 21cm maps from these simulations test the classifier's ability to discriminate between these models using information beyond the 21cm power spectrum.}
\label{fig:pk_match}
\end{figure}

In order to determine whether our classification algorithm is taking advantage of information in the 21cm images beyond what is available in the power spectrum, we consider a special case when the two models produce similar 21cm power. A Galaxies-only model yields 21cm power spectra at $z=7$ that are almost identical (at least in shape) to those obtained by the AGN-only model at $z=8$, within the neutral fraction range $0.45< x_{\rm HI} <0.95$. This is shown in Figure~\ref{fig:pk_match}. As the AGN-only models produce larger ionized bubbles that source a larger amplitude power spectrum, the Galaxies-only models require a greater contribution of power from the matter density field---available at lower redshifts due to structure growth---to compensate for this and to obtain a similar 21cm power spectrum. In any case, comparisons of different semi-numerical simulations reveal that different codes have better agreement when compared at matched neutral fractions rather than at matched redshifts~\citep{zah11}. This suggests considerable theoretical uncertainty in the ionization history, and thus it is not unreasonable to consider simulations shifted in redshift as we do here.

In artificially pairing boxes at different redshifts, it is important to also harmonize their respective noise simulations. The thermal noise depends on the system temperature ($\sqrt{\langle|N|^{2}\rangle} \propto T_\textrm{sys}$), which in turn depends on the redshift/frequency through the foreground sky model. Thus, adding a $z=7$ noise realization to our Galaxies boxes and a $z=8$ to our AGN boxes will result in different levels of signal to noise. As a result, the classifier might be able to discriminate between these models based on differences in $T_\textrm{sys}$. To exclude the possibility that the classifier doing this, we conservatively add the noise from $z=8$ to both two datasets.
\begin{figure}
\centering
\setlength{\epsfxsize}{0.5\textwidth}
\centerline{\includegraphics[scale=0.6]{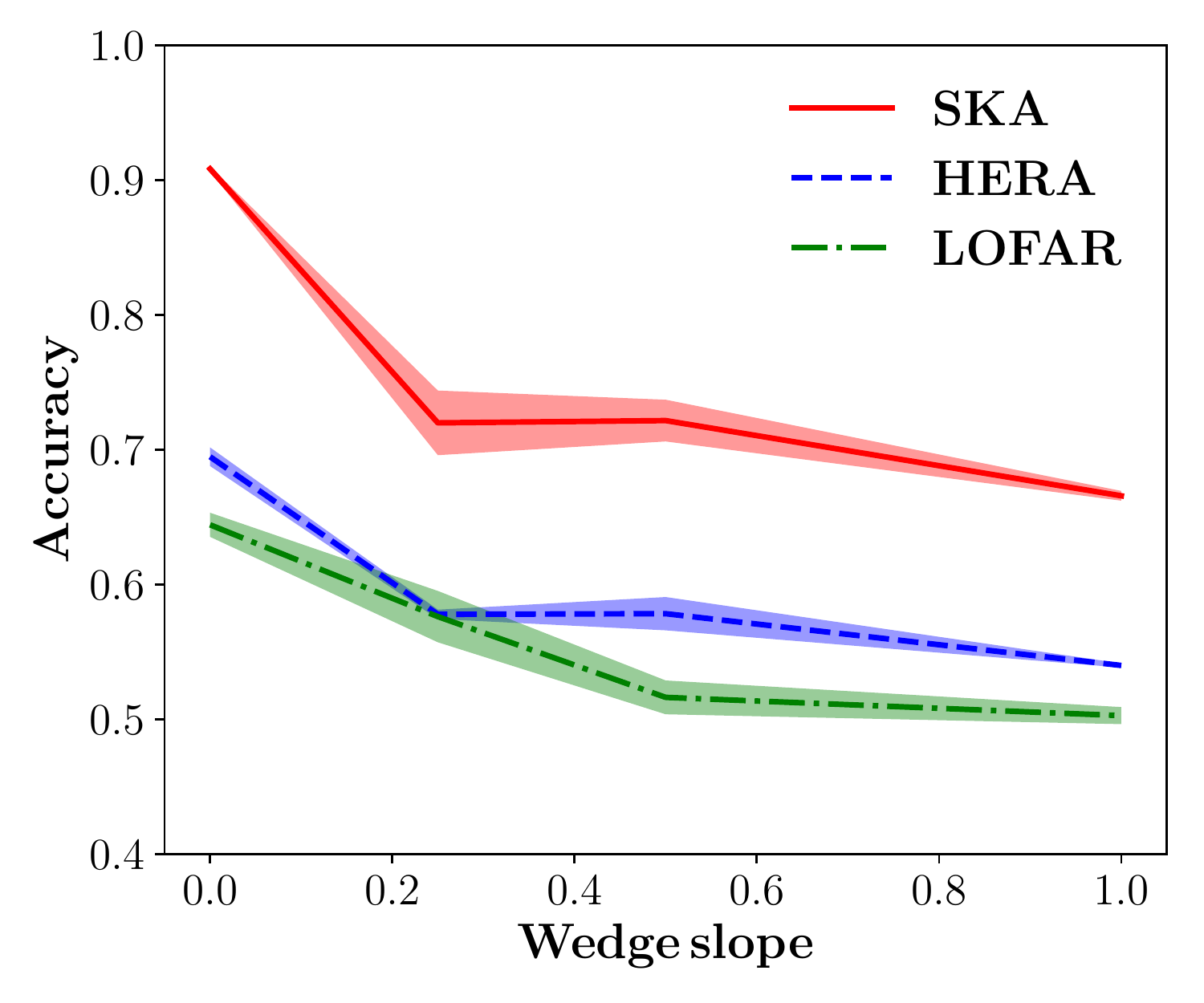}}
\caption{Same as Figure~\ref{fig:acc_noise} but for the case when the two models share similar power spectra (shown in Figure~\ref{fig:pk_match}). Fixing the power spectra of the two models makes it more difficult to distinguish between the models, and foregrounds must be 100\% successfully removed ($m=0.0$) for successful classification by the future SKA observations.}
\label{fig:acc_same_pk}
\end{figure}

We then re-do the same exercise as in the previous section and consider the effect of the angular resolution alone and that of the thermal noise with different wedge slopes with $m=0.0$, 0.25, 0.5, 1.0 to generate five different datasets for each of our experiments. We display the classifier accuracy as a function of the wedge slope in Figure~\ref{fig:acc_same_pk}. The effect of the angular resolution  here is much stronger than the previous case. The SKA is the only experiment is able distinguish between our models with an accuracy of $\sim$ 91\%. This is lower by a factor of $\sim$ 9\% as compared to the case without instrumental effects. At fixed power spectrum, neither HERA nor LOFAR is able to score an accuracy of more than 70\% for any level of thermal noise or foreground removal, since the angular resolution effects already present a significant challenge for these experiments in this case. Assuming 100\% successful foreground removal ($m=0.0$) and in the presence of thermal noise, the SKA only is able to discriminate between our models with a similar accuracy due to the angular resolution alone. However, all experiments fail to classify between our models if any amount of foreground contaminants is added.  Still, this shows that SKA is promising telescope for differentiating between our models (even with degenerate power spectra) when foreground removal is 100\% successful. We summarize the validation accuracy for the five different runs in Table~\ref{case3}. 

Unlike Figure~\ref{fig:acc_noise}, we find that the accuracy evolution as a function of wedge slope is not gradual. Here, we see that the accuracy from all experiments is almost constant between $m=1.0$ and $0.25$. This sets a limit of $m < 0.25$ for the SKA particularly to discriminate between Galaxies and AGN models. This limit is roughly a factor of four more stringent than the limit found in the previous section, suggesting that robust foreground mitigation is particularly important for breaking degeneracies that are present in a power spectral analysis alone.

\section{Conclusion}\label{sec:con}
In this work, we created a CNN-based classifier of reionization models that is able to discriminate between 21cm images produced by Galaxy- and AGN-driven reionization. In order to 
prepare this classifier for future 21cm surveys, we included instrumental effects for experiments such as the SKA, HERA, and LOFAR. We have studied the classifier performance as a function of redshift and neutral fraction. We further explored the impact of instrumental assumptions on the classification accuracy and the ability of these 21cm experiments to discriminate between our two models at fixed redshift and with power spectra tuned to be the same between the two scenarios.

Our key findings are as follows:
\begin{itemize}

\item The classifier accuracy increases at high redshifts and/or for restricted neutral fraction ranges. Excluding the case when the universe is highly ionized at $z=7$, the classifier is able to correctly classify more than 92-100\% of the images, with some dependence on redshift and neutral fraction range (see Figure~\ref{fig:zvsxHI} and Table~\ref{case1}).

\item At a fixed redshift of $z=8$ and neutral fraction range $0.25 < x_{\rm HI} < 0.75$, the SKA, HERA and LOFAR achieve $\sim$ 98\%, 95\% and 90\% accuracy in the presence of the thermal noise and the angular resolution effects but assuming 100\% successful foreground removal is achieved (no wedge is applied, $m=0.0$), respectively (see Figure~\ref{fig:acc_noise} and Table~\ref{case2}). The SKA is the only experiment is able to score an accuracy of $\gtrsim$ 90\% for wedge slope $m>0.0$ in this case.

\item If the power spectrum is constrained to be nearly identical between the two scenarios, the SKA is still able to achieve $\sim$ 91\% of accuracy in the presence of thermal noise and angular resolution effects, but again assuming 100\% successful foreground removal is achieved (see Figure~\ref{fig:acc_same_pk} and Table~\ref{case3}).

\item None of the experiments are able to achieve a classification accuracy of $\gtrsim90\%$ for conservative values of the wedge slope, which is about $m \sim 3-4$ at reionization redshifts (i.e. $z=6$ to $10$).

\end{itemize}
 
Our results show the preliminary promise of directly using 21cm images from future surveys to discriminate between Galaxy and AGN reionization scenarios. These direct constraints also have the potential to break degeneracies that would be present in a power spectrum analysis alone. For our proposed techniques to be maximally effective, however, it is important to be able to clean foreground contaminants beyond the most conservative wedge regions in Fourier space.

 While the foreground removal presents a significant challenge to our classifier, the CNN might still be able to perform perfectly well at a strong contamination levels if more careful post-processing to our training samples is applied. Such post-processing techniques include the use of super-pixels method as suggested by~\cite{giri18}, and the Principal Component Analysis to reduce the image dimensions and extract directly the 2-D information before passing the datasets to the CNN. Moreover, adding the redshift evolution information by using the light cones rather than static images might improve the CNN performance.  We leave exploring these possibilities for future works. 

It is worth mentioning that our results might be limited by the relatively small box size ($\sim$ 75 Mpc) and coarse resolution ($\sim$ 0.5 Mpc) used in this study. The differences between Galaxies and AGN reionization models might be more prominent for larger cosmological volumes and higher resolutions, hence relaxing the requirement for smaller wedge slopes. We leave exploring larger volumes and higher resolutions for future work. The approximation implemented in these semi-numerical models of Galaxies and AGN place an additional limitation to the presented results. Nonetheless, this work provides a step forward in our quest to extract astrophysical and cosmological information from future 21cm surveys to reveal the nature of cosmic reionization.

\section*{acknowledgements}
The authors acknowledges helpful discussions with J. Aguirre, B.A. Bassett, N. Gillet, S. Giri, A. Lidz, A. Mesinger, M. Molaro, J. Pober, and M. Santos.  We particularly thank the anonymous referee for the helpful and supportive comments that improved the paper significantly. SH acknowledges the financial assistance of the South Africa SKA Project (SKA SA) towards this research. SH also acknowledges the support received from CAMPARE-HERA Astronomy Minority Program (CHAMP/HERA), where part of this work was initially conducted at University of Pennsylvania.  AL acknowledges support for this work by NASA through Hubble Fellowship grant \# HST-HF2-51363.001-A awarded by the Space Telescope Science Institute, which is operated by the Association of Universities for Research in Astronomy, Inc., for NASA, under contract NAS5-26555. SAK is supported by a University of Pennsylvania SAS Dissertation Completion Fellowship.

\bsp	
\label{lastpage}
\end{document}